\documentclass[showpacs,preprintnumbers,amsmath,amssymb,floatfix]{revtex4-1}

\usepackage{color}      
\usepackage{graphicx}
\usepackage{subfig}
\usepackage{dcolumn}
\usepackage{bm}         

\usepackage{epsfig}

\begin{document}

\title{Pocket resonances in low-energy antineutrons reactions with nuclei}

\author{Teck-Ghee Lee$^1$}
\email{tgl0002@auburn.edu; tglee.physics@gmail.com}
\affiliation{$^1$Department of Physics, Auburn University, Auburn, Alabama 36849, USA.}

\author{Orhan Bayrak$^2$}
\email{bayrak@akdeniz.edu.tr}
\affiliation{$^2$Department of Physics, Akdeniz University, Antalya, Turkey}

\author{Cheuk-Yin Wong$^{3}$}
\email{wongc@ornl.gov}
\affiliation{$^3$Physics Division, Oak Ridge National Laboratory\footnote{
This manuscript has been authored in part by UT-Battelle, LLC, under
contract DE-AC05-00OR22725 with the US Department of Energy (DOE). The
US government retains and the publisher, by accepting the article for
publication, acknowledges that the US government retains a
nonexclusive, paid-up, irrevocable, worldwide license to publish or
reproduce the published form of this manuscript, or allow others to do
so, for US government purposes. DOE will provide public access to
these results of federally sponsored research in accordance with the
DOE Public Access Plan
(http://energy.gov/downloads/doe-public-access-plan), Oak Ridge,
Tennessee 37831, USA}, Oak Ridge, Tennessee 37831, USA.}

\begin{abstract}
Upon investigating whether the antineutron-nucleus annihilation
cross-sections at very low momenta $p$ satisfy Bethe-Landau's power
law of $\sigma_{\rm ann} (p) \propto 1/p^{\alpha}$, we uncover
unexpected regular oscillatory structures in the low antineutron
energy region from 0.001 to 10 MeV,  with small amplitudes and narrow
periodicity in the logarithm of the antineutron energies, for
large-$A$ nuclei such as Pb and Ag. Subsequent semiclassical analyses
of the $S$ matrices reveal that these oscillations are pocket
resonances that arise from quasi-bound states inside the effective
potential pocket with a barrier.  They are a continuation of the bound
states in the continuum.  Experimental observations of these pocket
resonances will provide vital information on the properties of the
optical model potentials and the nature of the antinucleon-nucleus reactions.
\end{abstract}

\maketitle

\section{Introduction}
One of the greatest mysteries in modern physics is the
matter-antimatter asymmetry in the Universe \cite{Sak67}.  To unravel
this mystery, there has been a great deal of experimental and
theoretical investigations on matter-antimatter interactions.  So far,
most of the obtained information centers around antiproton-nucleus
($\bar p$A) reactions and structures \cite{Ric20, asacusa18, Mau99,
  FAIR09, Bia11, Bal89, Kle05, Fri00, Fri01, Fri14, Lee14, Lee16,
  Lee08}.  The corresponding information on antineutron-nucleus ($\bar
n$A) reaction \cite{Ric20, Bres03}, on the other hand, remains comparatively
sparse and limited with the most recent work from the OBELIX
collaboration \cite{Bar97, Ast02} at CERN.  Nevertheless, the $\bar nA$
annihilation is essential in the process of quantifying signals from
$n\rightarrow\bar n$ oscillations in matter \cite{Vor20,Sno19, Lad19,
  Phi16}, and the significant connection between the $\bar n A$
interaction-potential and the $n \rightarrow \bar n$ oscillations
rates have also been examined theoretically \cite{Dov83, Kon96, Fri08}.
\begin{figure}[h]
\centering
\includegraphics[scale=0.39]{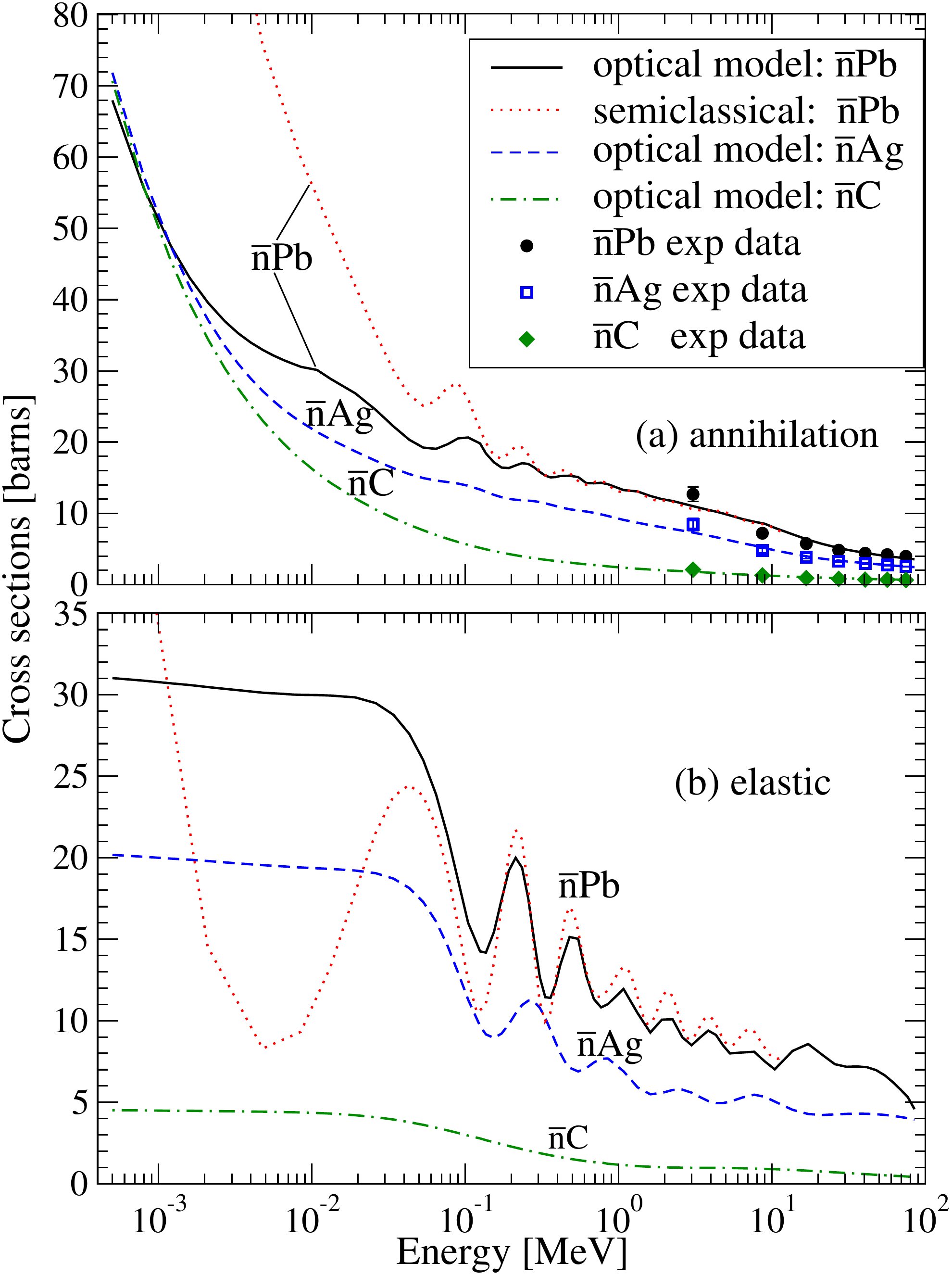} 
\caption{Optical-model cross sections as a function of energy.  (a)
  ${\bar n}A$ annihilation, (b) ${\bar n} A$ elastic scattering.  The red
  dotted-curve is the semiclassical result for the ${\bar n}$Pb
  reactions.}
\label{fig1}
\end{figure}

It has been generally expected that in the $s$-wave limit the $\bar n
A$ and $\bar p A$ annihilation cross-sections are to obey
Bethe-Landau’s power-law, $\sigma^{\bar n A, \bar p A}_{\rm ann}(p)
\propto 1/p^\alpha$ as a function of the antineutron momentum $p$,
with $\alpha=1$ for $\bar n A$ and $\alpha=2$ for $\bar p A$
\cite{Lan58}.  But recent experimental data (see Fig.\ 5 of
Ref.\cite{asacusa18}) revealed that the $\bar n$C and $\bar p$C annihilation 
cross-sections at low energies appear to follow
a similar trend with only minor differences. Much curious about the
puzzle, we recently introduced a new momentum-dependent optical
potential to investigate the behavior of the $\bar n$A, and the $\bar p$A annihilation
cross-sections on C, Al, Fe, Cu,
Ag, Sn, and Pb nuclei in the momentum range 50 to 500 MeV/$c$
\cite{Lee18}, via the ECIS code \cite{Ray81}. The calculated results
agree with the OBELIX's annihilation cross-section data
\cite{Bar97, Ast02}, but the optical model calculations indicated that
$\alpha \approx 1/2$ for $\bar n A$ and $\alpha \approx 1.5$ for $\bar
p A$ between 40 and 100 MeV/$c$, leading us to conclude 
that our low-energy annihilation reaction in question was yet to reach the $s$-wave limit.

While investigating the energy dependence of the cross-sections at
even lower energies, we uncover, to our surprise, unexpected regular
oscillatory structures with small amplitudes and narrow periods (in
the logarithm of the energy) in the annihilation and elastic
cross-sections, in the region from 0.001 to 10 MeV, as shown in
Fig.~\ref{fig1}. Such oscillations are absent for small nuclei and gain
in strength as the nuclear radius increases. Its amplitude is larger
for the elastic scattering than for the annihilation process.  Such
behavior, undoubtedly, contradicts Bethe-Landau’s power-law. It is
reminiscent of a potential resonance behavior.  However, the predicted
oscillations appear somewhat different from that of neutron-induced
total cross sections for Pb, Cd, and Ho nuclei, which arises
from the Ramsauer's resonances \cite{Ram21, Pet62, Mar70} that are
broad potential resonances without a barrier whose peak position moves
to higher energies as the nuclear mass increases. They are also unlike
those present in the high-energy $^{12}$C+$^{12}$C fusion, which is
due to successive addition of contributions from even values of
partial waves to the identical-particle fusion cross-section, with
increasing energies \cite{Esb12,CYW12,Row15}.

To understand the nature of these oscillations, we carry out a
semiclassical $S$ matrix analysis for $\bar n $Pb reaction in which the
oscillations are quite prominent.  As shown in Fig.~\ref{fig1} 
the structure of the cross-section oscillations in the semiclassical 
calculations for $\bar n $Pb agrees approximately 
with the structure obtained in the optical model calculations, 
except for the lowest energy region for which the semiclassical approximation is not applicable. 
We wish to show that the cross-section oscillations are physical and constitute
narrow pocket resonances that can be best described as arising from
quasi-bound states inside an effective potential pocket with a
barrier, as depicted in Fig.\ 5 of Ford and Wheeler \cite{FW59}. Such
pocket resonances are expected to be common features of potential
scattering in nuclear, atomic, molecular, and heavy-ion collisions because they
are the continuation of the bound states in the continuum.  They move
to lower energies as the mass number of the nucleus increases as is
shown in Fig.~\ref{fig1}.  To understand the origin of such resonances,
a thorough analysis of the reaction process in detail is therefore
needed in what follows.

\section{Semiclassical analyses of the oscillations of the annihilation and elastic cross sections}

Semiclassical descriptions of the potential scattering problem have
been discussed in great details by Ford and Wheeler \cite{FW59} using
the Wentzel-Kramers-Brillouin (WKB) approximation, and later
consolidated by other pioneering works \cite{Mil68, BM72, BT77, Lee78,
  Ohk87}. However, we shall follow the Brink-Takigawa formulation
\cite{BT77} for complex-potential scattering that was also applied by
Lee, Marty, and Ohkubo \cite{BT77, Lee78, Ohk87}.
\vspace*{-0.8cm}
\begin{figure}[h]
\centering \includegraphics[scale=0.32]{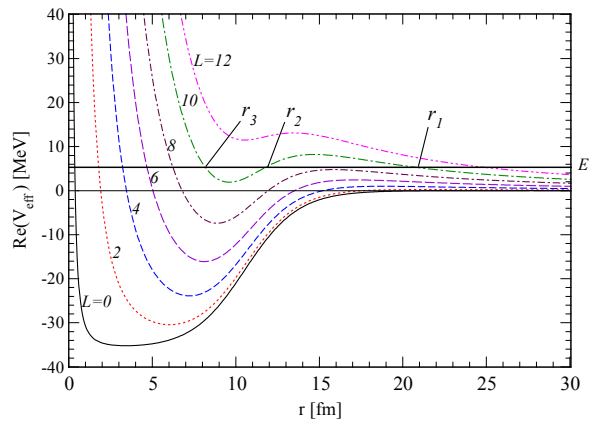}
\caption{ Real part of the $\bar n$Pb interaction potential Re($V_{\rm
    eff}(r,E)$) at a sample energy $E$, where only the potential
  curves for even orbital angular momentum $L$ are displayed.  The
  $r_{1,2,3}$ denote three turning points for Re($V_{\rm eff}(r,E)$).}
\label{fig2}
\end{figure}
 Accordingly, we consider the interference between a barrier wave
 reflected at the potential barrier and modified by the tunneling effect, 
 and an internal wave that penetrates the barrier, into the
 potential pocket, and reemerges through the barrier out to $r\to
 \infty$.

For a given $E$ and a partial wave $L$, the total phase shift
$\delta=\delta(E,L)$ is related to the WKB phase function $\delta_{\rm
  WKB}$, the action phase angle $\delta_{ij}$ between the turning
points $r_i$ and $r_j$, tunneling phase $\phi$, tunneling phase angle
$\xi$ and tunneling coefficient $w$ by \cite{BT77, Lee78, Ohk87},
\begin{eqnarray}
\hspace*{-1.0cm}
&&\hspace*{-1.0cm}\delta=\delta_{\rm WKB} - \phi 
+ \tan^{-1}\{w(\xi)\tan(\delta_{32}-\phi)\},
\\
&&\hspace*{-1.0cm}\delta_{\rm WKB}=\frac{\pi}{2}(L+\frac{1}{2}) - kr_1 + \int_{r_1}^{\infty} (K-k) dr, ~~~~~~
\\
\label{wkbphase}
&&\hspace*{-1.0cm}\delta_{ij} = \int_{r_i}^{r_j}Kdr,
\label{act} 
\\
&&\hspace*{-1.0cm} \phi =  \frac{1}{2}\left(\xi \ln(\xi/e) - \arg \Gamma(\frac{1}{2} + i\xi) \right),\\
 \label{tunnelphase} 
 &&\hspace*{-1.0cm}\xi =  -\frac{i}{\pi}\delta_{21},
\label{tunnelcoef} 
\\
 &&\hspace*{-1.0cm}w(\xi) = \frac{\sqrt{(1+e^{-2\pi\xi})}-1}{\sqrt{(1+e^{-2\pi\xi})}+1},
 \label{wcoef}  
\end{eqnarray}
where $K$ = $\sqrt{2\mu[E-V_{\rm eff}(r,E)]/\hbar^2}$, $k = \sqrt{2\mu E/\hbar^2}$, $E$ is
the center-of-mass energy, and $\mu$ is the reduced mass of the
collision pair.  The turning points $r_{1,2,3}$ are the roots of
$E-V_{\rm eff}(r,E)=0$, as displayed in Fig.~\ref{fig2}. In our
case, $V_{\rm eff}(r,E)$ is the Langer's modified interaction
potential,
\begin{eqnarray}
\hspace*{-0.4cm}
V_{\rm eff}(r,E)  = \frac{-(V(E)+iW)}{1+\exp\{(r-R)/a_o\}}+\frac{\hbar^2(L+1/2)^2}{2\mu r^2}
\label{Vpot}.
\end{eqnarray} 
We adapt the same momentum-dependent optical potential used in our
earlier work \cite{Lee18} in the present semiclassical theory.  The
values for the parameters $r_o=R/A^{1/3}$, $a_o$ and volume terms $V(E)$
and $W$ are listed in Table 1, 2, and 3 in \cite{Lee18}. To keep the
analyses simple, we consider only the nuclear volume terms in the
present semiclassical theory.

Figure \ref{fig2} illustrates the real part of the potential
Re($V_{\rm eff}(r,E))$ as a function of internuclear distance $r$ and
different $L$ at a sample energy $E$ for the $\bar n$Pb
interaction. With the nuclear potential that has a negative imaginary
potential $W$, the turning points $r_{1,3}$ are below the real axis of
the complex-plane and $r_2$ lies above, and the $V_{\rm eff}(r, E)$ for
a given $E$ must be an analytic function of $r$. The integrands in the
action phase angle $\delta_{ij}$ are multiple-valued quantities involving branch cuts. 
The integration paths in the complex plane need to be chosen so that the 
action phase angles $\delta_{ij}$ and $\xi$ are positive, 
as discussed in \cite{BT77}.

As mentioned earlier in the Introduction, the semiclassical cross-sections 
for $\bar n$Pb reaction give an approximate representation of the optical model results,
 except for the lowest energy region.  Our task is to carry out a detailed 
 semiclassical partial wave analysis to pinpoint the origin of the cross-section oscillations.

So we concentrate on the interference between
the barrier waves and the internal waves and split the total $S$
matrix element into the barrier wave contribution $S_B$, and the
internal wave contribution $S_I$:
\begin{eqnarray}      
S&=&  e^{2i\delta }= S_B + S_I,    \label{eq8} \\
S_B &=& \frac{e^{2i\delta_{\rm WKB}}}{N}, 
\label{smatBI1}   \\ 
S_I &=& \left[\frac{e^{2i\delta_{\rm WKB}}}{N}\right]e^{-2\pi\xi} 
 \left[\frac{e^{2i\delta_{\rm 32}}}{N}\right] \nonumber \\
 && \times\left[1+\frac{e^{2i\delta_{\rm 32}}}{N}+\left(\frac{e^{2i\delta_{\rm 32}}}{N}\right)^2+\left(\frac{e^{2i\delta_{\rm 32}}}{N}\right)^3
 +\cdots\right] \nonumber \\
&=& \left[\frac{e^{2i\delta_{\rm WKB}}}{N}\right]e^{-2\pi\xi}\left[\frac{e^{2i\delta_{\rm 32}}}{N}\right]
       \left[\frac{1}{1+e^{2i\delta_{\rm 32}}/N}\right],
\label{smatBI2}   
\end{eqnarray}
where $N=\sqrt{2\pi} \exp(-2\pi \xi+i\xi
\ln(\xi/e))/\Gamma(\frac{1}{2}+i\xi)$ is the barrier-penetration
factor.  The factor $1/(1+e^{2i\delta_{32}}/N)$ in the $S_I$ term is
from Pad\'{e} approximation, where $e^{2i\delta_{32}}/N$ is the
amplitude of penetration-weighted internal-wave.  Thus, the factor
$e^{2i\delta_{32}}/(N+e^{2i\delta_{32}})$ essentially describes the
sum of amplitudes of multiple reflections of internal-wave inside the
potential pocket \cite{BT77,Lee78}. Knowing these $S$-matrices, $\sigma^{\rm tot}_{\rm ann}
= \sum_L \sigma_{\rm ann}(L) = ({\pi}/{k^2})\sum_L(2L+1)(1-|S|^2)$ 
and  $\sigma^{\rm tot}_{\rm el} = \sum_L\sigma_{\rm el}(L) = 
 ({\pi}/{k^2})\sum_L(2L+1)(|1-S|^2)$ define the cross sections 
 for annihilation and elastic scattering, can be readily evaluated.

Partial wave analysis of the annihilation oscillations in terms of the 
interference of the barrier and the internal waves gives \cite{Ohk87}
\begin{eqnarray}  
 \hspace*{-1.0cm}   && \sigma _{\rm ann}^{\rm tot}=\Sigma^{\rm BI}_{\rm ann}
+\Sigma^{\rm int}_{\rm ann}=
\sum\limits_{L=0}^{\infty} (\sigma^{\rm BI}_{\rm ann}(L) + \sigma^{\rm int}_{\rm ann}(L) ),
\label{ann1} \\
 \hspace*{-1.0cm}  &&  \sigma^{\rm BI}_{\rm ann}(L)=\frac{\pi}{k^2} (2L+1) (1-|S_B|^2-|S_I|^2),
\label{annBI}  \\
  \hspace*{-1.0cm} &&  \sigma^{\rm int}_{\rm ann}(L)=\frac{\pi}{k^2} (2L+1) (-2{\rm Re}\{S_BS^{\ast}_I\}), 
\label{annint} 
\end{eqnarray}
where $\sigma_{\rm ann}^{\rm int}(L)$ arises from the interference between the
internal and barrier waves, and $\sigma_{\rm ann}^{\rm BI}(L)$ from the
remaining non-interference term.
 
\begin{figure}[!tbp]
\centering
\includegraphics[scale=0.35]{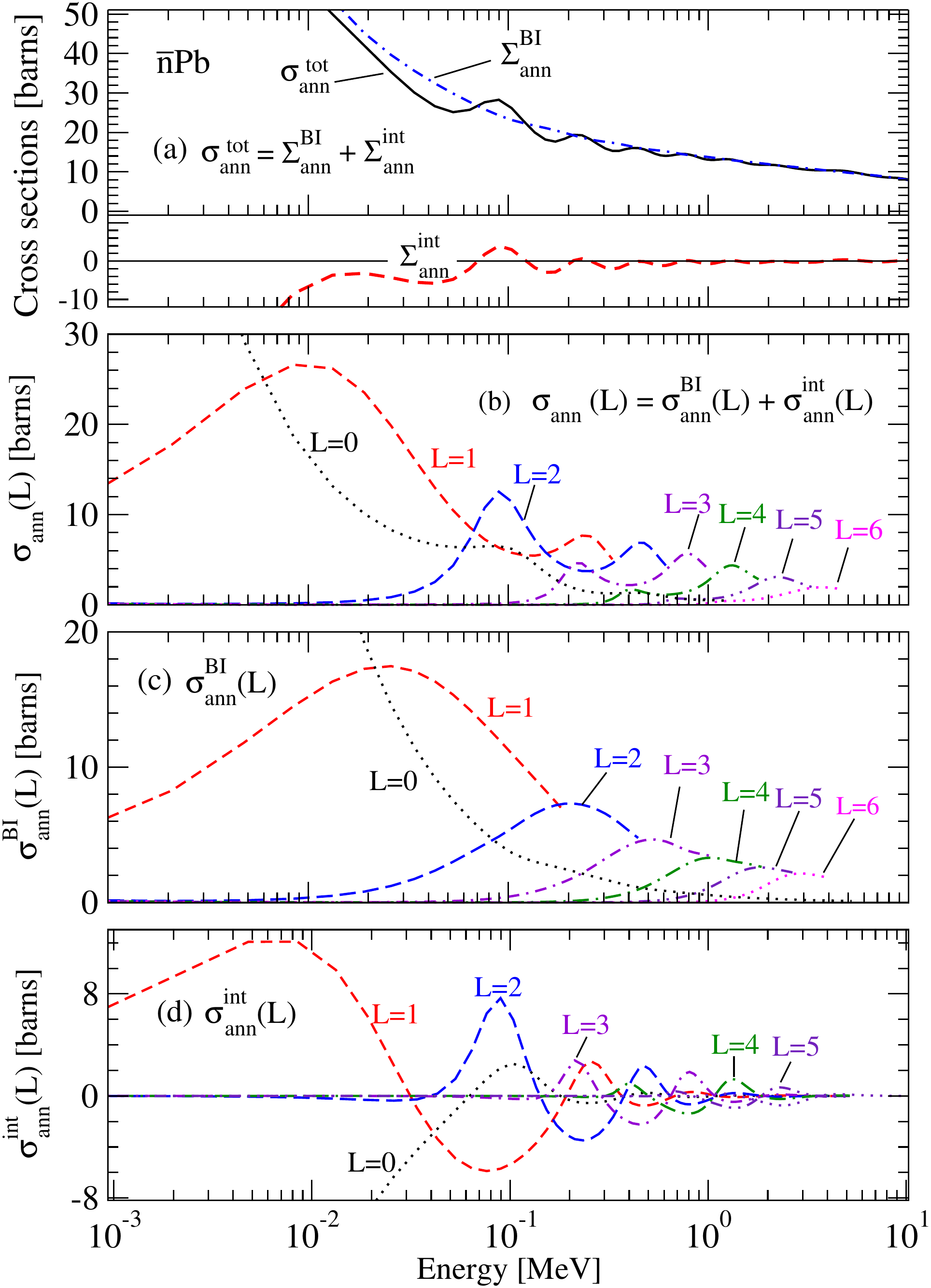}
\caption{ Semiclassical results for $\bar n$Pb annihilation as
  a function of energy.  (a) is the total cross section of Eq.(\ref{ann1}), 
  (b) is the partial-wave cross section of Eq.(\ref{ann1}),
  (c) is the non-interference cross section of Eq.(\ref{annBI}) and 
  (d) is the interference cross section of Eq.(\ref{annint}) }.
\label{fig3}
\end{figure}

We show in Fig.~\ref{fig3}(a) the results of the separation of the total
annihilation cross section $ \sigma _{\rm ann}^{\rm tot}$ of Eq.\  (\ref{ann1}) into the
non-interference part, $\Sigma^{\rm BI}_{\rm ann}$, and the
interference part, $ \Sigma^{\rm int}_{\rm ann}$ as a function of $E$. 
The non-interference $\Sigma^{\rm BI}_{\rm ann}$ produces the familiar
$1/p$-like dependence with decreasing energy without any oscillations.
In contrast, the interference part $\Sigma^{\rm int}_{\rm ann}$ does oscillate, 
 giving rise to the total (sum of all partial-wave) 
annihilation cross section $\sigma^{\rm tot}_{\rm ann}$ that oscillates on the smooth
$1/p$ background as shown in the plot.

Figure \ref{fig3}(b) shows the decomposition of the total annihilation cross
section, $\sigma^{\rm tot}_{\rm ann}$, into different partial waves contributions,
$\sigma_{\rm ann}(L)$. Except for the $L$ = 0 
partial-wave that resembles the $1/p$-like behavior, the other partial waves 
display a double peak. Fig.~\ref{fig3}(c) shows the $\sigma_{\rm ann}^{\rm BI}(L)$ 
for different $L$ as given by Eq.(\ref{annBI}), indicating a monotonically 
decreasing cross section for $L=0$ and a single broad peak for higher partial waves. 

\begin{figure}[t]
\centering
\includegraphics[scale=0.35]{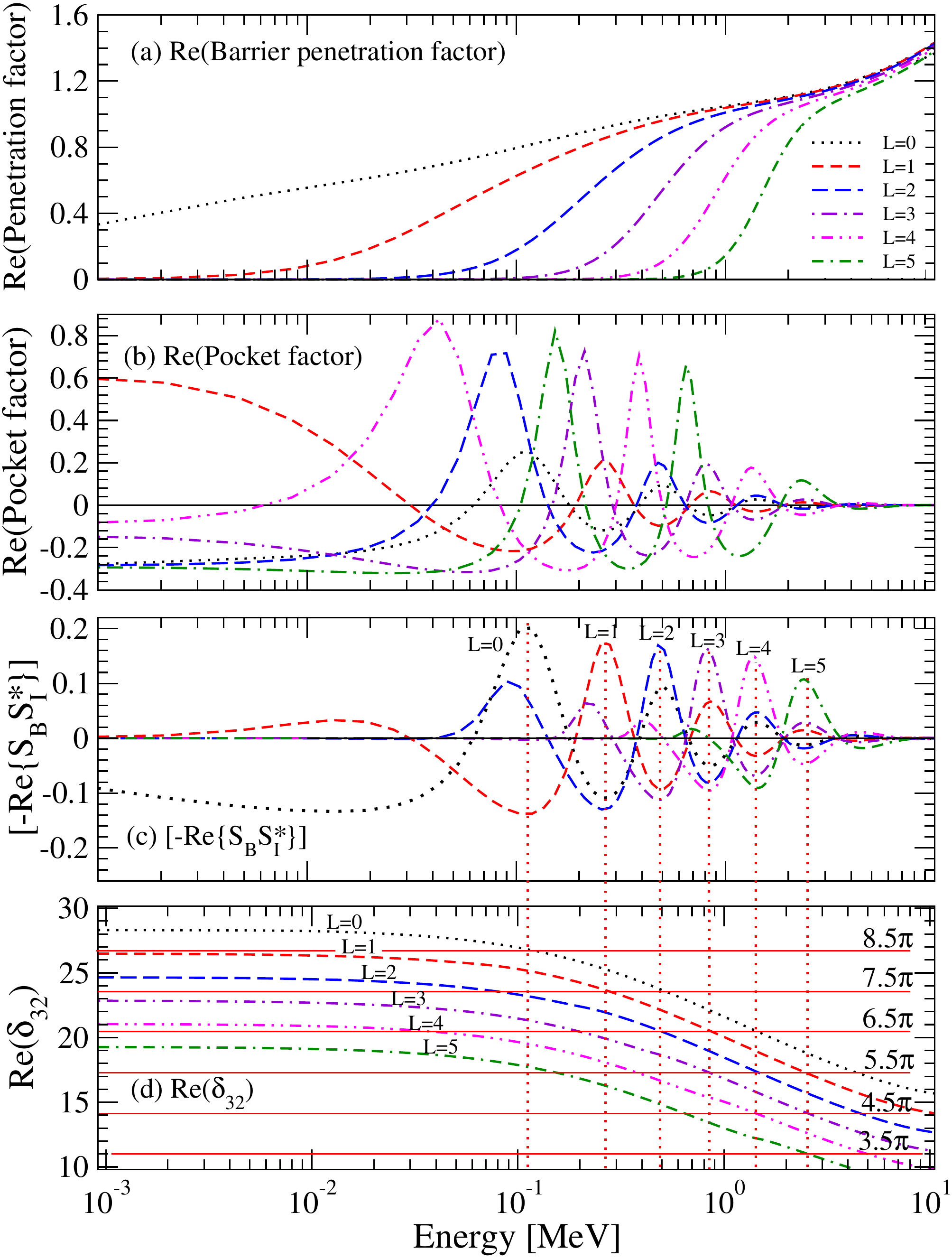}
\caption{For different partial
  waves $L$, the energy dependence of 
  (a) the real part of the barrier penetration  factor in Eq.\ (\ref{SISB}), 
   (b) the real part of the pocket factor in Eq.\ (\ref{SISB}), (c) the quantity $[-{\rm Re}\{S_BS^{\ast}_I\}]$, and (d) the quantity Re($\delta_{3 2}$). }
\label{fig4}
\end{figure}

Figure \ref{fig3}(d) displays 
$\sigma_{\rm ann}^{\rm int}(L)$ for different $L$ as
given by Eq.(\ref{annint}).    Inspection of Eq.(\ref{smatBI1}),
(\ref{smatBI2}) and (\ref{annint}) reveals that the oscillations of $\sigma_{\rm ann}^{\rm int}(L)$ are
governed by $[-2{\rm Re}(S_BS^{\ast}_I)]$ in which $[-(S_BS^{\ast}_I)]$ can be
factorized into
\begin{equation}    
\hspace*{-0.4cm}-S_BS^{\ast}_I\!=\hspace*{-0.1cm}
\underbrace{
\left[\frac{e^{2i\delta_{\rm WKB}}}{N}\right]
\left[\frac{e^{2i(\delta_{\rm WKB}+\delta_{\rm 21})}}{N}\right]^{\ast}
}_{\rm barrier~penetration~factor}
\underbrace{
\left[\frac{-e^{2i\delta_{\rm 32}}/N}{1+e^{2i\delta_{\rm 32}}/N}\right]^*
}_{\rm pocket~factor}\!, 
\label{SISB}
\end{equation}
which depends on $\delta_{\rm WKB}$, $\delta_{2 1}$, and $\delta_{3 2}$. 
The real part of the above  barrier penetration factor and  
pocket factor in  Eq.(\ref{SISB}) for various $L$
are displayed in Fig.~\ref{fig4} as a function of $E$.  In the
$r_2<r<r_1$ region below the barrier, $\delta_{2 1}$ are predominantly
imaginary, resulting in the pentration probability $\exp(2i\delta_{21}) \approx \exp(-2\pi\xi)$
with $\xi>0$.  The ($\delta_{\rm WKB} - \delta^*_{\rm WKB})$ difference
of the $r>$ $r_1$ outer-most
region are also predominantly imaginary, 
rendering $\exp[2i(\delta_{\rm WKB}-\delta^*_{\rm WKB})] \approx
\exp(-2\eta)$ with $\eta>0$. We can therefore recast Eq.(\ref{annint}) into
\begin{equation}    
\hspace*{-0.3cm}\sigma^{\rm int}_{\rm ann} (L)\! \sim\! \frac{\pi}{k^2} (2L\!+\!1)|S_B||S_I| e^{-2(\pi\xi+\eta)}[ -2\cos (2\delta_{3 2})].\!
\label{sigintcos}
\end{equation}

Consequently we find that the maxima of the annihilation cross
sections $\sigma_{\rm ann}^{\rm int}(L)$  in Fig.~\ref{fig3}(d) as
``annihilation resonances" are located at the maxima of the real part
of the pocket factor of Eq. (14) in Fig.~\ref{fig4}(b) and the maxima of  $[-{\rm Re}(S_BS_I^*)]$ 
in Fig.~\ref{fig4}(c), as marked by the vertical dashed lines
in Fig.~\ref{fig4}(c) and \ref{fig4}(d) for some of the resonances. 
Even though the barrier penetration factor of Fig.~\ref{fig4}(a) 
increases approximately stepwise with increasing energy and modifies the
oscillation amplitude of the pocket factor of Fig.~\ref{fig4}(b), the final positions 
of the oscillation maxima as shown in Fig.~\ref{fig4}(c) remain
approximately unchanged. 
The maxima of the pocket factor, according to
Eq.(\ref{sigintcos}), occur whenever $\cos (2\delta_{3 2})=-1$
(or $2\delta_{3 2}=(2n+1)\pi$).  The subsequent maxima 
of $[-{\rm Re}\{S_BS^{\ast}_I\}]$ are indeed located at
$\delta_{3 2} \approx$ 8.5$\pi$, 7.5$\pi$ and 4.5$\pi$, etc., respectively,
as indicated by horizontal lines in Fig.~\ref{fig4}(d). The quantity
$\delta_{32}$ varies by one $\pi$ unit between resonances  of the same $L$ with
$\partial\delta_{3 2}/\partial n \approx -\pi$.  We further find that
wherever the horizontal lines intersect the $\delta_{3 2}$ curves of
Fig.~\ref{fig4}(d), each intersection point approximately, and respectively, 
matches each maximum of $[-{\rm Re}\{S_BS^{\ast}_I\}]$ in Fig.~\ref{fig4}(c) and each maximum of $\sigma_{\rm ann}^{\rm int}(L)$ 
in Fig.~\ref{fig3}(d).   This means that the condition for an
annihilation resonance at a cross-section maximum at the energy $E$
is
\begin{eqnarray}
\delta_{3 2}(E,L)  \approx (n_L-n_L' +\frac{1}{2})\pi,
\label{quan}
\end{eqnarray}
where $n_L$ is the number of bound and quasi-bound radial 
states lower than $E$ for the angular momentum $L$ 
and 
$n_L'$ (= 0, 1, 2, $\cdots$) is the additional resonance number above the first resonance of the same $L$.  
The above sign of $n_L'$ in Eq.\ (\ref{quan}) depends on $\partial V/\partial E$ as a function of $E$.  

Since $\delta_{32}(E,L)$ is the action phase angle integrating over the confining
potential pocket of $V_{\rm eff}(r,E))$, condition of Eq.(\ref{quan}) 
is just the quantization condition for a quasi-bound state
inside the pocket.  Thus, the ``pocket resonance" is,
in fact, a continuation of the bound states in the continuum trapped in a pocket with a barrier as depicted in Fig.\ 5 of Ford and Wheeler
\cite{FW59}.  The condition of Eq.\ (\ref{quan}) for the resonances suggests
that pocket resonances occur quite generally in potential scattering
problems.  Their observations however depend on the degree of absorption as we shall discuss later.

\begin{figure}[h]
\centering
\includegraphics[scale=0.35]{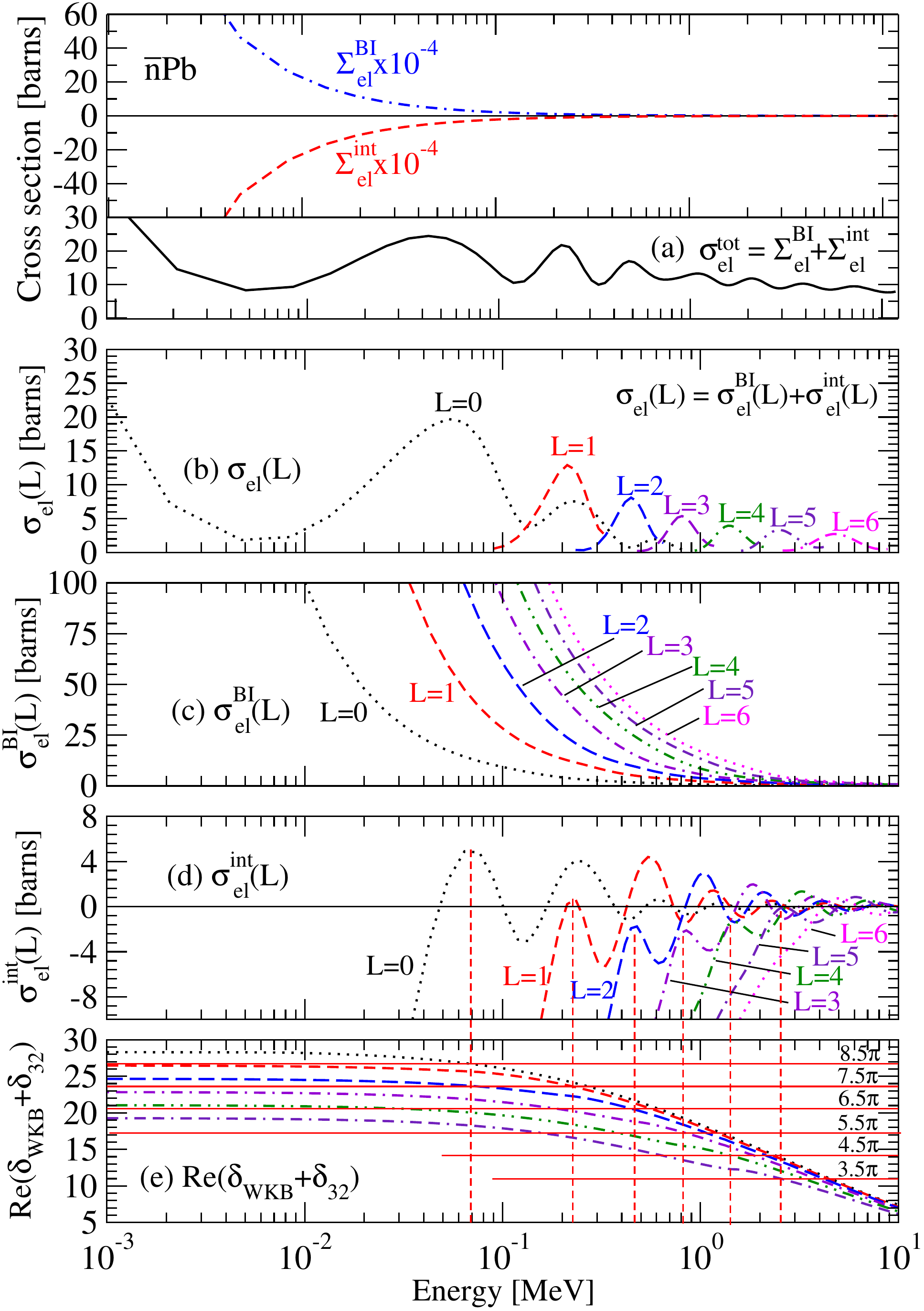}
\caption{ Semiclassical results for $\bar n$Pb elastic scattering as a
  function of energy.  (a) is the total cross section of Eq.(\ref{elas1}), 
  (b) is the partial-wave cross section of Eq.(\ref{elas1}) for different $L$, (c) is the
  non-interference cross section of Eq.(\ref{elasBI}), (d) is the interference cross section of Eq.(\ref{elasint}) and (e) is 
  the Re$(\delta_{\rm WKB} + \delta_{32})$ for different $L$.}
\label{fig5}
\end{figure}

We now examine the elastic oscillations with     
\begin{eqnarray}      
&&\hspace*{-0.5cm} \sigma _{\rm el}^{\rm tot}=\Sigma_{\rm el}^{\rm BI}+\Sigma_{\rm el}^{\rm int}=\sum\limits_{L=0}^{\infty}(\sigma^{\rm BI}_{\rm el}(L) + \sigma^{\rm int}_{\rm el}(L)) ,
\label{elas1} \\
&&\hspace*{-0.5cm}\sigma^{\rm BI}_{\rm el}(L)=\frac{\pi}{k^2}{(2L + 1)(1+|S_I|^2+|S_B|^2)}, 
\label{elasBI}  \\
&&\hspace*{-0.5cm} \sigma^{\rm int}_{\rm el}(L)=  \frac{\pi}{k^2}{(2L + 1)(2{\rm Re}\{S_I^*S_B-S_I^*-S_B\}}).
\label{elasint}  
\end{eqnarray} 

Figure \ref{fig5}(a) indicates that there is a significant cancellation of
$\Sigma_{\rm el}^{\rm BI}$ by $\Sigma_{\rm el}^{\rm int}$, where
$\Sigma_{\rm el}^{\rm int}$ has an oscillating behavior, resulting in
an oscillating total elastic cross section $\sigma_{\rm el}^{\rm tot}$ as a
function of energy as shown in Fig.~\ref{fig5}(a).

Figure \ref{fig5}(b) shows the decomposition of the total elastic
$\sigma^{\rm tot}_{\rm el}$ into different partial waves contributions
$\sigma_{\rm el}(L)$.  Only the $L$ = 0 partial wave has three,
narrow, sinusoidal oscillations with decreasing amplitudes. The $L >$
0 curves, however, have only a single maximum with a similar
shape. The first, broad maximum seen in $\sigma^{\rm tot}_{\rm el}$ near 0.05
MeV closely corresponds to the first maximum of $L$ = 0 curve at the
same energy.  The subsequent, second maximum of $\sigma^{\rm tot}_{\rm el}$ that
is near 0.2 MeV corresponds to the contributions from the $L$ = 1
curve and the second maximum of $L$ = 0 curve. Beyond the second
maximum, each consecutive maximum now appears to be associated with a
specific partial wave of $L$ = 2, 3, 4, and so forth.
Fig.~\ref{fig5}(c) shows that $\sigma^{\rm BI}_{\rm el}(L)$ for
different $L$ from Eq.(\ref{elasBI}) decrease monotonically as energy
increases.

Figure \ref{fig5}(d) is more interesting as it exhibits 
the oscillating $\sigma^{\rm int}_{\rm el}(L)$ as a
function of $E$ for different $L$.  The oscillations are similar to
those in the annihilation case.  Although Eq.(\ref{elasint}) is more
complicated than Eq.(\ref{annint}) for the annihilation, we have found
from our numerical calculations that the $[-2{\rm Re}{(S_I)}]$ term
dominates in the evaluation of $\sigma_{\rm el}^{\rm int}(L)$ over the
energy range.  Furthermore, the maximum of $\sigma^{\rm int}_{\rm el}(L)$ as an
``elastic resonance" for various partial waves is located at the same
energy as the maximum of $[-{2\rm Re}{(S_I)}]$.  The phase of $S_I$, 
according to Eq.(\ref{smatBI2}), is given predominantly by $(\delta_{\rm WKB}+\delta_{32})$.  
Numerical calculations of $(\delta_{\rm WKB}+\delta_{32})$ indicate 
that the difference of the phases between elastic resonances is $\pi$, 
i.e., $\partial (\delta_{\rm WKB}+\delta_{32})/\partial n \approx -\pi$. 
  
Together with Fig.~\ref{fig5}(d), the plot of \ref{fig5}(e) shows the condition 
for the elastic resonances at the energy $E$ arising from the pocket is
\begin{eqnarray}
(\delta_{\rm WKB} + \delta_{32})(E,L) \approx (n_L-n_L'+\frac{1}{2})\pi,
\end{eqnarray} at which 
$\cos[2(\delta_{\rm WKB} + \delta_{32})(E,L)]=-1$, substantiating the
quantization condition for the elastic pocket resonance similar to
that for the annihilation pocket resonance in Eq.(\ref{quan}), but
with the addition of $\delta_{\rm WKB}$ to $\delta_{32}$. And because
of this additional $\delta_{\rm WKB}$ phase the location of the elastic
pocket resonance energy $E$ is slightly shifted from the annihilation
pocket resonance energy for the same $n$ and $L$.

Another interesting feature for both the annihilation and elastic
resonances is that the resonance energies shift toward lower
energies with increasing nuclear radius or mass, as is clear from
Fig.~\ref{fig1}(b)  in comparing $\bar n$Ag and $\bar n $Pb elastic cross-sections.  
This arises because the pocket resonances are the continuation of the bound states in the
continuum. Note that the opposite behavior occurs where the Ramsauer
resonance shifts toward higher energies with increasing nuclear radius
in the neutron total cross-section data for Cu, Cd, and Pb
\cite{Bow61} and Ho \cite{Mar70} nuclei from 2 MeV to 125 MeV. The
cause of the broad resonance and its shifts to higher energies have
been explained by Peterson \cite{Pet62} and Marshak $et.~al.$ \cite {Mar70}
using the concept of Ramsauer's interference \cite{Ram21}.

As the incident energy further decreases below 0.03 MeV, we see the
optical model predicted annihilation cross-sections for C, Ag, and Pb
nuclei merge into one single curve. Parametrizing these curves with
$\sigma^{\bar nA}_{\rm ann} (p) \propto 1/p^{\alpha_{\bar N}}$ in this
energy region (i.e., 1.0 $\le p \le$ 2.0 MeV/$c$), we found that
$\alpha_{\bar N} = \partial \ln(\sigma^{\bar nA}_{\rm ann})/\partial
\ln(p)$ for C, Ag and Pb nuclei to be $\sim$1.0, $\sim$0.9, and
$\sim$0.8, respectively, which are surely close to Bethe-Landau's $s$-wave
prediction of $\alpha_{\bar n}$ = 1.  On the elastic scattering at
energies below 0.03 MeV, we notice that the scattering becomes isotropic
and energy-independent. Consequently, the ``$s$-wave" cross-section is
purely geometrical effect and a constant, which can be described by a
black-nucleus model with $\sigma_{\rm el}^{\rm tot}=\pi R^2$ \cite{Bla52}.

It is important to point out the sensitivity of the
pocket resonances on the optical model parameters $W$, $V$, and $R$ of Eq. (\ref{Vpot}).  
For the pocket resonances to exist, the lifetime of the incident particle inside
the potential pocket, $\hbar /2W$, must be longer than the passage
time for the particle to traverse from the entrance to the exit in
the pocket, of order $2R/v$, where $v$ is the particle velocity in
the pocket. This corresponds to the requirement of 
$W$ to be approximately less than $W_o \sim
\hbar \sqrt{2(E+|V|)/m_{\bar n}}/4R$. For the $\bar n$Pb reaction, 
$W_o \sim $ 2.3 MeV (estimated) for potential parameters extrapolated 
to the pocket resonance energy region.  

The available $\bar n$Pb annihilation cross-section
data are consistent with the value of $W=2.8$ MeV \cite{Lee18}, which is
comparable to the above estimate of onset value  $W_o$.  Hence, the
pocket resonances are slightly diminished in oscillatory magnitudes
but remain to be present for the $\bar n$Pb reaction in
extrapolation, as shown in Fig.~\ref{fig1}. Other choices of $W$ and
radius parameter would not lead to a good description of the
experimental annihilation data. The oscillatory amplitudes of pocket
resonances will be much sharper for very small values of $W$ but
decrease as $W$ increases, and the oscillatory pocket resonances
will be absent if $W$ is substantially large.

It has been known from the works of Brink and Takigawa
\cite{BT77, Lee78} that there is an anomalous large-angle scattering
(ALAS) elastic scattering that is sensitive to the imaginary part $W$
of the optical potential. The elastic scattering at large angles at a
pocket resonance requires the passage of the incident particle from
the entrance to the exit of the pocket and is likewise sensitive to
the degree of absorption represented by $W$. We find that as a pocket
resonance is associated with a specific angular momentum $L$ which
shows up as a dip in a particular angle in the elastic angular
distribution at large angles, the shape of the dip, and the swing of
the angular distribution at larger scattering angles is sensitively
affected by the strength of $W$. There is also the ALAS phenomenon
associated with the pocket resonance as energy is varied across the
resonances energy region, similar to the other ALAS observed in
Ref.\cite{BT77,Lee78} .

\vspace*{-0.2cm}
\section{Conclusions}
Optical-model calculations show oscillatory structures in $\bar nA$
annihilation and elastic cross-sections in the low $\bar n$ energy
range 0.001 to 10 MeV.  The cross-sections oscillate in the logarithm 
of the antineutron energies with small amplitudes and narrow periods 
for large nuclei. This surprising behavior contradicts the generally expected
Bethe-Landau's power-law in the $s$-wave limit.

Semiclassical $S$-matrix analysis provides new insight 
into the nature and the origin of the oscillations.  Two important contributions to
the structures in the reaction cross sections are found: (1) For
annihilation and elastic scattering, the oscillations are attributed
to the interference between the internal and barrier waves, while
their smooth, $1/p$, backgrounds are mainly due to the
non-interference term. (2) Delving deeper into the interference term
for both reactions, we identified the maxima of the cross-sections as
resonances occurred within the potential pocket.  The condition for an
annihilation resonance is the quantization rule: $\delta_{3 2}(E,L)
\approx (n_L-n_L'+1/2)\pi $, and for an elastic resonance: $(\delta_{\rm
  WKB}+\delta_{3 2})(E,L) \approx (n_L-n_L' +1/2)\pi $ where $n_L$ is the
the number of bound and quasi-bound radial states in the pocket below $E$ for the partial wave $L$,
and $n_L'$ is the additional resonance number.
So the existence of resonances is connected to the potential pocket
and barrier that depends sensitively on optical potential parameters.
Experimental observations of these pocket resonances, if they occur,
will provide valuable information on the properties of the optical
model potentials and the nature of the reaction process. While here we have
focused on $\bar nA$ reactions, similar potential pocket resonances 
are also expected for reactions involving the use of
an optical potential in nuclear, atomic, molecular, and heavy-ion
collisions such as in $^{12}$C+$^{12}$C reactions  \cite{Tang20}, which is an
important reactions in nuclear astrophysics.

\vspace{-0.1in}
\section*{Acknowledgment}
\vspace{-0.1in}
CYW's research is supported in part by the Division of Nuclear Physics, 
U.S. Department of Energy under Contract DE-AC05-00OR22725.


\vspace{-0.1in}
\begin{thebibliography}{9}

\bibitem{Sak67}
A. D. Sakharov,  JETP. Lett. {\bf 5} 24  (1967).


\bibitem{Ric20} 
J-M. Richard., Front. Phys. {\bf 8}, 1 (2020).

\bibitem{asacusa18}
H.Aghai-Khozani $et~ al.$, Nucl. Phys. A {\bf 970},  366  (2018); Hyperfine Interact. {\bf 234},  85  (2015); Hyperfine Interact. {\bf 229},  31  (2014).

\bibitem{Mau99} S. Maury, (for the AD Team), {\it the Antiproton
Decelerator (AD)}, CERN/PS 99-50 (HP) (1999).

\bibitem{FAIR09}
 {\it FAIR
- Facility for Antiproton and Ion Research}, Green Paper, October 2009. 


%
%
%
%
\bibitem{Bia11} 
A. Bianconi, $et~al.$, Phys. Lett. B {\bf 704} 461 (2011); 
%
A. Bianconi, $et~al.$, Phys. Lett. B {\bf 481}, 194 (2000);
%
A. Bianconi, $et~al.$, Phys. Lett. B {\bf 492}, 254  (2000).


\bibitem{Bal89}
F. Balestra, $et~al.$, Phys. Lett. B {\bf 230}, 36  (1989);
%
F. Balestra, $et~al.$, Nucl. Phys. A {\bf 452}, 573 (1986);
%
F. Balestra, $et~al.$, Phys. Lett. B {\bf 165}, 265 (1985);
F. Balestra, $et~al.$, Phys. Lett. B {\bf 149},  69 (1984).



\bibitem{Kle05}
E. Klempt, C. Batty and J.-M. Richard,
Phys. Rep. {\bf 413},  197 (2005).

\bibitem{Fri00} 
A. Gal, E. Friedman, and C. J. Batty, Phys. Lett. B {\bf 491}, 219 (2000).
 
\bibitem{Fri01} 
C. J. Batty, E. Friedman, and A. Gal, Nucl. Phys. A. {\bf 689}, 721 (2001).

\bibitem{Fri14}
E. Friedman, Nucl. Phys. A {\bf 925}, 141 (2014); Hyperfine Interact. {\bf 234}, 77 (2015).

\bibitem{Lee14}
T. G. Lee and C. Y.  Wong, Phys. Rev. C {\bf 89}, 054601 (2014).

\bibitem{Lee16}
T. G. Lee and C. Y.  Wong, Phys. Rev. C {\bf 93}, 014616 (2016).

\bibitem{Lee08}
T. G. Lee, C. Y. Wong, and L. S. Wang, Chin. Phys. {\bf 17} 2897 (2008);
C. Y. Wong and T. G. Lee, Ann. Phys. {\bf 326}, 2138 (2011). 

\bibitem{Bres03}
T. Bressani and A. Filippi, Phys. Rep. {\bf 383} 213 (2003).




\bibitem{Bar97}
C. Barbina $et~ al.$, Nucl. Phys. A {\bf 612},  346  (1997).


\bibitem{Ast02} 
M. Astrua, $et~al.$, Nucl. Phys. A {\bf 697}, 209 (2002).  


\bibitem{Vor20}
K. V. Protasov, V. Gudkov, E. A. Kupriyanova, V. V. Nesvizhevsky, W. M. Snow, and A. Yu. Voronin., 
Phys. Rev. D., {\bf 102}, 075025 (2020).

\bibitem{Sno19}
V. V. Nesvizhevsky, V. Gudkov, K. V. Protasov, W. M. Snow, and A. Yu. Voronin., 
Phys. Rev. Lett., {\bf 122}, 221802 (2019). 

\bibitem{Lad19}
E. S. Golubeva, J. L. Barrow, and C. G. Ladd, 
Phys. Rev. D., {\bf 99}, 035002 (2019) 

\bibitem{Phi16}
D. G. Phillips $et~ al.$, Physics Reports {\bf 612}, 1 (2016). 

\bibitem{Dov83}
C. B. Dover, A. Gal, and J. M. Richard, Phys. Rev. D {\bf 27},
1090 (1983); Phys. Rev. C {\bf 31}, 1423 (1985); Nucl. Instrum.
Methods Phys. Res., Sect. A {\bf 284}, 13 (1989).

\bibitem{Kon96}
L. A. Kondratyuk., 
JETP. Lett. {\bf 64}, 495 (1996).

\bibitem{Fri08}
E. Friedman and A. Gal., Phys. Rev. D {\bf 78}, 016002 (2008). 

\bibitem{Lan58}
L. D. Landau and E. M. Lifshitz, {\it Quantum Mechanics} (Pergamon, Oxford 1958).


\bibitem{Lee18}
T. G. Lee and C. Y.  Wong, Phys. Rev. C {\bf 97}, 054617 (2018).

\bibitem{Ray81} 
ECIS97~(https://people.nscl.msu.edu/$\sim$brown/reaction-codes \\
/home.html); J. Raynal., Phys. Rev. C., {\bf 23}, 2571 (1981)
 
\bibitem{Ram21}
C. Ramsauer, Ann. Phys., {\bf 66}, 546 (1921).  

\bibitem{Pet62}
J. M. Peterson, Phys. Rev. {\bf 125}, 955 (1962). 

\bibitem{Mar70}
H. Marshak, A. Langford, C. Y. Wong, and T. Tamura, Phys. Rev. Lett., {\bf 20}, 554 (1968);  
H. Marshak, A. Langford, T. Tamura and C. Y. Wong, Phys. Rev. C {\bf 5}, 1862 (1970).

\bibitem{Esb12}
H. Esbensen, Phys. Rev. C {\bf 85}, 064611 (2012).

\bibitem{CYW12}
 C. Y. Wong, Phys. Rev. C {\bf86}, 064603 (2012).

\bibitem{Row15}
N. Rowley and K. Hagino, Phys. Rev. C {\bf 91}, 044617 (2015).


\bibitem{FW59}
K. W. Ford and J. A. Wheeler, Ann. Phys., {\bf 7}, 239 (1959); {\bf 7}, 259 (1959).

\bibitem{BM72}
M. V. Berry and K. L Mount, Rep. Prog. Phys. {\bf 35} 315 (1972).

\bibitem{Mil68}
W. H. Miller, Adv. Chem. Phys. {\bf 25}, 69 (1974).

\bibitem{BT77}
D. M. Brink and N. Takigawa, Nuc. Phys. A {\bf 279} 159 (1977).

\bibitem{Lee78}
S. Y. Lee, N. Takigawa, and C. Marty, Nuc. Phys. A {\bf 308} 161 (1978).

\bibitem{Ohk87}
S. Ohkubo and D. M. Brink, Phys. Rev. C {\bf 36}, 966 (1987).


\bibitem{Bow61}
P. H. Bowen, J. P. Scanlon, G. H. Stafford, J. J. Thresher, P. E. Hodgson., Nucl. Phys. {\bf 22}, 640 (1961).





\bibitem{Bla52}
J. M. Blatt and V. F. Weisskopf, {\it Theoretical Nuclear Physics}, John Wiley and Sons, N.Y., 1952, p. 324.

\bibitem{Tang20}
N. T. Zhang {\it et al.}, Physics Letters B, {\bf 801}, 135170 (2020).

\end{thebibliography}
\end{document}